\documentclass{aa520}
\usepackage{txfonts}
\usepackage{graphicx}
\usepackage{natbib}
\usepackage[figuresright]{rotating}
\bibpunct{(}{)}{;}{a}{}{,}
\begin{document}

\title{The age of the Galactic thin disk from Th/Eu
nucleocosmochronology}

\subtitle{II. Chronological analysis}

\titlerunning{The age of the Galactic thin disk from Th/Eu
nucleocosmochronology. II}

\author{E.F. del Peloso\inst{1} \and L. da Silva\inst{1} \and
L.I. Arany-Prado\inst{2}}

\offprints{E.F. del Peloso}

\institute{Observat\'orio Nacional/MCT, Rua General Jos\'e
Cristino
77, 20921-400 Rio de Janeiro, Brazil\\
\email{epeloso@on.br, licio@on.br} \and Observat\'orio do
Valongo/UFRJ, Ladeira
do Pedro Ant\^onio 43, 20080-090 Rio de Janeiro, Brazil\\
\email{lilia@ov.ufrj.br}}

\date{Received <date> / Accepted <date>}

\abstract{The purpose of this work is to resume investigation of
Galactic \emph{thin disk} dating using nucleocosmochronology with
Th/Eu stellar abundance ratios, a theme absent from the literature
since 1990. [Th/Eu] abundance ratios for a sample of 20 disk
dwarfs/subgiants of F5 to G8 spectral type with
$-0.8\le\mathrm{[Fe/H]}\le+0.3$, determined in the first paper of
this series, were adopted for this analysis. We developed a
Galactic chemical evolution model that includes the effect of
refuse, which are composed of stellar remnants (white dwarfs,
neutron stars and black holes) and low-mass stellar formation
residues (terrestrial planets, comets, etc.), contributing to a
better fit to observational constraints. Two Galactic disk ages
were estimated, by comparing literature data on Th/Eu production
and solar abundance ratios to the model
($8.7^{+5.8}_{-4.1}~\mbox{Gyr}$), and by comparing [Th/Eu] vs.
[Fe/H] curves from the model to our stellar abundance ratio data
(($8.2\pm1.9)~\mbox{Gyr}$), yielding the final, average value
$(8.3\pm1.8)~\mbox{Gyr}$. This is the first Galactic disk age
determined via Th/Eu nucleocosmochronology, and corroborates the
most recent white dwarf ages determined via cooling sequence
calculations, which indicate a low age ($\lesssim10\mbox{ Gyr}$)
for the disk.

 }

\maketitle

\section{Introduction}

Current estimates of the age of the Galactic thin
disk\footnote{All references to the \emph{Galactic disk} must be
regarded, in this work, as references to the \emph{thin} disk,
unless otherwise specified.} are obtained by dating the oldest
open clusters or white dwarfs. Ages of open clusters are
determined by fitting isochrones, and ages of white dwarfs are
determined using cooling sequences. Isochrone and cooling sequence
calculations require deep knowledge of stellar evolution, are very
complex, and depend on a large number of physical parameters known
at different levels of uncertainty. Many important aspects of
stellar evolution, like the influence of rotation, are not well
known, and may have a strong influence on the outcome of isochrone
calculations. Furthermore, white dwarf cooling sequences also
depend on calculations of degenerate-matter physics.
Nucleocosmochronology is a dating method that makes use of only a
few results of main sequence stellar evolution models, therefore
allowing a quasi-independent verification of the afore mentioned
techniques.

Nucleocosmochronology estimates timescales for astrophysical
objects and events by using abundances of radioactive nuclides.
These are compared with the abundances of their daughter nuclides,
or of other nuclides that are created by the same or a similar
nucleosynthetic process. Depending on the half-life of the chosen
nuclide, different timescales can be probed. The Th/Eu
chronometer, first proposed by \citet{pagel89}, is adequate to
assess the age of the Galactic disk, since \element[][232]Th is a
radioactive nuclide with a 14.05~Gyr half-life (i.e., of the order
of magnitude of the age being assessed). Eu is a satisfactory
element for comparison, because it is produced almost exclusively
\citep[97\%, according to\space][]{burrisetal00} by the same
nucleosynthetic process that produces all Th, the rapid
neutron-capture process (r-process).

Since \citet{snedenetal96} carried out the first Th/Eu
nucleocosmochronological dating of an ultra-metal-poor (UMP) star,
the literature has been virtually dominated by these objects
\citep[\space and references therein]{truranetal02}. The reason
for such interest lies in the strong simplifications that can be
applied to the dating of UMP stars, which does not require the use
of Galactic chemical evolution (GCE) models. Dating of disk stars,
on the other hand, does require the use of GCE models, and is thus
a very intricate process. Owing to the complexity of the analysis,
\citet{dasilvaetal90} is the only work ever published for disk
stars, presenting preliminary results for only four objects. This
work aims at resuming investigation of Galactic disk dating using
[Th/Eu] abundance ratios.\footnote{In this paper we employ the
following customary spectroscopic notations: absolute abundance
$\log\varepsilon(\mbox{A})\equiv\log_{10}(N_{\mathrm{A}}/N_{\mathrm{H}})+12.0$,
and abundance ratio
$\mbox{[A/B]}\equiv\log_{10}(N_{\mathrm{A}}/N_{\mathrm{B}})_{\mathrm{star}}
-\log_{10}(N_{\mathrm{A}}/N_{\mathrm{B}})_{\mbox{\scriptsize\sun}}$,
where $N_{\mathrm{A}}$ and $N_{\mathrm{B}}$ are the abundances of
elements A and B, respectively, in atoms~cm$^{-3}$.}

In the first paper of this series \citep[\space
Paper~I]{delpelosoetal05a}, [Th/Eu] abundance ratios were
determined for a sample of 20~disk dwarfs/subgiants of F5 to G8
spectral type with $-0.8\le\mathrm{[Fe/H]}\le+0.3$. In what
follows, we employ a GCE model developed by us to determine two
Galactic disk estimates: one by comparing Th/Eu production and
solar abundance ratios from the literature with the model, and one
by comparing the stellar abundance data determined by us with
[Th/Eu] vs. [Fe/H] curves obtained from the model.

\section{Nucleocosmochronological analysis}

Stars that compose our sample have been formed all along the disk
lifetime. Th present in them was synthesized in a number of
stellar generations, the younger stars receiving contributions
from a greater number of them; consequently, matter synthesized by
each generation has decayed for different amounts of time. On that
account, a GCE model is indispensable for a correct interpretation
of the evolution of the [Th/Eu] abundance ratios, from which we
derive the Galactic disk age.

\subsection{Galactic chemical evolution model}

\subsubsection{Basic description}

We have developed a GCE model based on \citet[\space
PT95]{pagel&tautvaisiene95} with the inclusion of the effect of
refuse, following an ameliorated version of the formulation of
\citet{rochapintoetal94}. The inclusion of the effects of refuse,
which are composed of stellar remnants (white dwarfs, neutron
stars, and black holes) and low-mass stellar formation residues
(terrestrial planets, comets, etc. --  hereafter referred to as
simply \emph{residues}), contribute to a better fit of the Fe
abundance evolution model to the metallicity distribution of the G
dwarfs in the solar neighbourhood.

The model of PT95 is composed of multiple phases. According to
\citet{beers&sommerlarsen95}, about 30\% of the stars with
$\mbox{[Fe/H]}\le-1.5$ within 1~kpc of the Galactic plane belong
to the disk, which shows a considerable intersection between the
disk and halo metallicity distributions. Several other works
reveal this intersection (PT95, and references therein). Moreover,
\citet{wyse&gilmore92} argue that the angular momentum
distribution function of the halo stars makes it highly improbable
that the gas removed from the halo during its stellar formation
phase was ever captured by the disk, but more likely by the bulge.
Hence, disk and halo initiated their formations disconnected from
each other, both from the primordial, unenriched gas, and thus the
disk can be modelled independently from the halo. Our model is not
rigorously a \emph{Galactic} chemical evolution model, but rather
a Galactic \emph{disk} chemical evolution (GDCE) model.
Accordingly, we assume a simple model
\citep{vandenbergh62,schmidt63} until $\sim0.03T_{\mathrm{G}}$,
where $T_{\mathrm{G}}$ is the Galactic disk age. After this
initial phase, we assume the model of \citet{clayton85}, with a
mass infall rate
\begin{equation}
F=\omega g\frac{k}{u+u_0},
\end{equation}
where $\omega$ is a constant that represents the efficiency of
interstellar gas conversion into stars, $g$ is the mass of
interstellar gas, $u=\omega t$ is a dimensionless time-like
variable, $u_0$ is an arbitrary parameter, and $k$ is a small
positive integer. We define a time-like variable $u_1$, which
corresponds to the transition between simple and Clayton models.

We adopted the delayed production approximation developed by
\citet{pagel89}, which takes into account the delay in production
of elements that are synthesized mainly in stars of slow
evolution, like Fe, which is predominantly generated in Type~Ia
supernovae (with a typical timescale of 1~Gyr). In this
approximation, it is assumed that the elements whose production is
delayed start being ejected at a time~$\Delta$ after the start of
stellar formation. The abundance $A_i$ of element $i$ is comprised
of two components: a prompt component $A_{i1}$ and a delayed
component $A_{i2}$, so that $A_i=A_{i1}+A_{i2}$. Some elements,
like Th, are generated exclusively by prompt processes, and
consequently have a null delayed component ($A_{i2}=0$). Others,
like Fe, are formed by both kinds of processes, and thus have
non-null prompt and delayed components.

The original PT95 model was modified as follows. In
\citet{rochapintoetal94}, it is assumed that residues evaporate a
considerable amount of H and He, retaining metals and diluting the
interstellar medium (ISM). This dilution effect is equivalent,
mathematically, to a second source of metal-poor infall, which is
important for GDCE models \citep{chiappinietal97}. In our
calculations, the parameter~$\Gamma$ represents the dilution of
the ISM due to residue evaporation, and is defined as
\begin{equation}
\label{eqn:gamma}\Gamma=\frac{(\mbox{Jovian planet and residue
formation rate})}{(\mbox{stellar formation rate})-(\mbox{stellar
ejection rate})},
\end{equation}
where the stellar formation includes the formation of brown dwarfs
and the ejection accounts for the returned fraction (to the ISM)
due to stellar death.

\begin{figure}
\resizebox{\hsize}{!}{\includegraphics*{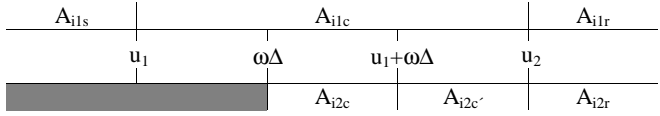}}
\caption{Structure of the adopted GDCE model (not to scale). $A_i$
is the abundance of element~$i$. The indices 1 and 2 represent the
prompt and delayed contributions, respectively. The indices $s$,
$c$, $c^{\prime}$, and $r$ are relative to the simple model ($s$),
Clayton model with delayed production contribution from stars born
during the simple model phase ($c$) and after the start of infall
($c^{\prime}$), and ISM dilution due to effect of refuse ($r$).}
\label{fig:model_phases}
\end{figure}

We assumed that residue formation, and hence the effect of refuse,
becomes efficient when metallicity becomes high enough to allow
the coagulation of plan\-e\-tes\-i\-mals
($\mbox{[Fe/H]}\sim-0.30$). The time-like variable $u_2$, which
cor\-res\-ponds to this moment, was set initially at
$u_2/\omega=0.40~T_{\mathrm{G}}$, but was later revised to
$0.42~T_{\mathrm{G}}$ by fitting observational constraints. The
final structure of our model is made up of five phases:
\begin{enumerate}\setlength{\itemsep}{0mm}
\item $u<u_1$: simple model, \item $u_1\le u<\omega\Delta$:
Clayton model with instantaneous recycling, \item $\omega\Delta\le
u<u_1+\omega\Delta$: Clayton model with delayed production
contribution from stars born during the simple model phase, \item
$u_1+\omega\Delta\le u<u_2$: Clayton model with delayed production
contribution from stars born after the start of infall, \item
$u\ge u_2$: ISM dilution due to effect of refuse.
\end{enumerate}
A schema depicting these phases is presented in
Fig.~\ref{fig:model_phases}.

Eu evolution is modeled by the same formulation used for Fe. Th
requires specific formulation due to its radioactive decay.
Another important phenomenon, besides radioactive decay, must be
taken into consideration when modeling Th evolution: selective
destruction by photonuclear reactions in the stellar interior
\citep{malaneyetal89}. During H burning via the CN cycle, 7.29~MeV
and 7.55~MeV $\gamma$~photons are emitted. These photons have
energies below the photonuclear threshold of most elements
existing in stellar interiors, including Fe and Eu. However, Th
has a threshold of $\sim6~\mbox{MeV}$, which allows the
destruction of up to $\sim40\%$ of its initial content in a
stellar core. When the star evolves to become a red giant, the
first dredge-up takes Th-depleted matter from the stellar interior
to the photosphere. Matter ejected at the end of stellar evolution
has a reduction in Th abundance of $\sim10\%\mbox{--}20\%$. This
phenomenon is properly considered in \citet[\space
APM98]{aranyprado&maciel98}, following \citet{malaneyetal89}. The
general equation for the abundance $A_i$ is
\begin{equation}
\frac{\mathrm{d}(gA_i)}{\mathrm{d}u}+(J_1+J_2)gA_i=P_i,
\end{equation}
where
\begin{equation}
J_1 = \left\{ \begin{array} {l @{ \quad} l l}
1, & \mbox{for }u & <u_2, \\
1+\Gamma, & \mbox{for }u & \ge u_2,
\end{array} \right.
\end{equation}
\begin{equation}
J_2 = \left\{ \begin{array} {l @{ \quad} l l}
0, & \mbox{for stable elements,} \\
\frac{\lambda_i}{\omega}+\Lambda_i, & \mbox{for r-process
radionuclides,}
\end{array} \right.
\end{equation}
\begin{equation}
P_i = \left\{ \begin{array} {l l @{ \quad} l l}
p^i_1g(u), & \hspace{-0.15cm} &\mbox{for }u & \hspace{-2.1cm}<\omega\Delta,\mbox{\ or for} \\
& & \mbox{prompt-enrichment}&\\
& & \mbox{elements}, \mbox{and}&\\
p^i_1g(u)+p^i_2g(u-\omega\Delta), & \hspace{-0.15cm} & \mbox{for
}u & \hspace{-2.1cm}\ge \omega\Delta.
\end{array} \right.
\end{equation}
$\lambda_i$ is the decay constant and $\Lambda_i$ is the
destruction term (APM98, Eq.~9) of element~$i$; $p^i_1$ and
$p^i_2$ are the net yields of element~$i$, as defined by Eq.~8 of
APM98 (as well as by Eqs.~6 and 7 of PT95); the indices~1 and 2
refer to the prompt and delayed contributions, respectively.

\subsubsection{Observational constraints and adopted constants}

Calculations were carried out for four Galactic disk ages: 6, 9,
12, and 15~Gyr. Parameter $u_1$ was kept at the value assumed by
PT95 (0.14). The delay $\Delta$ was fixed at 1.1~Gyr both for Fe
and for Eu; this value is consistent with the average evolution
timescales of progenitors of Type~Ia supernovae (which are the
main astrophysical sites of Fe production) and asymptotic giant
branch stars (which are the sites of s-process Eu production).
Note that delayed production of Eu is properly accounted for in
this work, but that its effects are nonetheless small, since Eu is
majorly \citep[97\%, according to\space][]{burrisetal00}
synthesized by the r-process.

We adopted the initial mass function (IMF) of \citet[\space
Eq.~6]{kroupa01}, which covers non-stellar masses (brown dwarfs)
down to $0.010~m_{\mbox{\scriptsize\sun}}$. We extended this lower
segment down to $0.001~m_{\mbox{\scriptsize\sun}}$, keeping the
same slope, and included a new segment corresponding to the
residues that evaporate, with a slope~$j$ which is a free model
parameter. The complete IMF was defined as follows:
\begin{equation}
\label{eqn:imf} \Phi(m) = \left\{ \begin{array} {l @{, \quad} l l}
a_1\,m^{-j} & \mbox{for }0.000 & \lesssim m/m_{\mbox{\scriptsize\sun}}\le\mathrm{\ \ }0.001; \\
a_2\,m^{-0.3} & \mbox{for }0.001 & <m/m_{\mbox{\scriptsize\sun}}\le\mathrm{\ \ }0.080; \\
a_3\,m^{-1.8} & \mbox{for\ }0.080 & <m/m_{\mbox{\scriptsize\sun}}\le\mathrm{\ \ }0.500; \\
a_4\,m^{-2.7} & \mbox{for }0.500 &
<m/m_{\mbox{\scriptsize\sun}}\le\mathrm{\ \ }1.000; \\
a_5\,m^{-2.3} & \mbox{for }1.000 &
<m/m_{\mbox{\scriptsize\sun}}\le62.000.
\end{array} \right.
\end{equation}
It is worth noting that the IMF for the residues relates to the
mass of residues prior to evaporation. If the IMF from
\citet{miller&scalo79} had been used, the slope of the residues
mass range would have to be steeper to compensate for the decrease
at low mass range, maintaining normalisation. The sudden change of
slope between residue and brown dwarf ranges would be harder to
explain than the slower, more continuous increase that can be
achieved when using \citeauthor{kroupa01}'s IMF. The same would
occur if we adopted the IMF from \citet{salpeter55}.

The parameters $\omega$ and $u_0$ were evaluated for different
values of $k$ and for each Galactic disk age, so that at
$T_{\mathrm{G}}$ we have $\mu$ = present-day gas fraction = 0.11
and $g$ = mass of gas ${\sim(0.73\,\mbox{initial disk mass})}$, as
considered by PT95. For a given set of values of the parameters
$k$, $\omega$, $u_o$, and T$_{\rm G}$, we fitted the metallicity
distribution of G dwarfs obtained from the model to the
observational distribution of \citet{rochapinto&maciel96}. The
fitting was accomplished by changing the parameter $j$ and,
consequently, the parameter $\Gamma$, since the Jovian planet and
residue formation rate is given by
$\int_0^{0.010}m\Phi(m)\,\mathrm{d}m$ (cf.
Equations~\ref{eqn:gamma} and \ref{eqn:imf}). For a given value of
parameter~$j$, parameters $a_1$, $a_2$, $a_3$, $a_4$, and $a_5$
(Equation~\ref{eqn:imf}) were obtained by enforcing IMF continuity
and normalisation. The best values of $j$ (2.562--2.605) are
similar to the IMF slopes for the ranges of stellar masses that
would form planets $(2.7 \mbox{\ for\ }
0.5<m/m_{\mbox{\scriptsize\sun}}\le1.0 \mbox{\ and\ } 2.3 \mbox{\
for\ } 1.0<m/m_{\mbox{\scriptsize\sun}}\le62.0)$, which implies
that equal masses form stars and residues -- see
Table~\ref{tab:GDCE_model_parameters}. In this table,
$f_1=p_1/A_{\mbox{\scriptsize\sun}}$ and
$f_2=p_2/A_{\mbox{\scriptsize\sun}}$.

\begin{figure}
\resizebox{\hsize}{!}{\includegraphics*{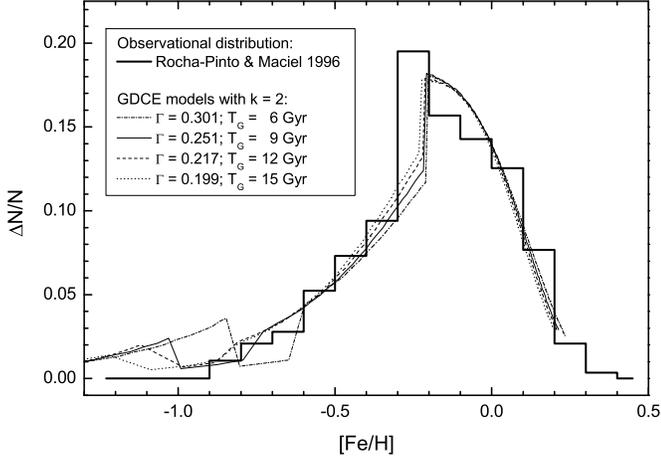}}
\caption{Metallicity distribution of G dwarfs in the solar
neighbourhood, where $\Delta\mathrm{N}/\mathrm{N}$ is the
normalised number of stars. The thick curve is the observational
distribution of \protect\citet{rochapinto&maciel96}. The thin
curves were calculated using our GDCE model for the four different
Galactic disk ages shown in the legend.} \label{fig:fe_dmd}
\end{figure}

Good fits were found for $k=2$ and $k=3$; we adopted $k=2$,
because this value provides a better fit to the four Galactic disk
ages simultaneously. This lends support to the hypothesis that the
effect of refuse can substitute for the dilution caused by infall
at later stages of Galactic evolution, since PT95 manage to obtain
good fits only with $k=3$. The final metallicity distributions of
G dwarfs obtained from the model are compared to
\citet{rochapinto&maciel96} in Fig.~\ref{fig:fe_dmd}.

For the modeling of Eu and Th evolutions, the relevant parameters
are their respective $f_1$ and $f_2$. For Eu, the total value
$f_{\mathrm{total\,Eu}}$, which takes prompt and delayed
production into consideration, was determined by fitting the
[Eu/Fe] vs. [Fe/H] curve obtained from the calculations to our
stellar data (Table~14 of Paper~I). We assumed that prompt
production of Eu is accomplished by the r-process, while delayed
production is carried out by the s-process (with relative
contributions 0.97 and 0.03, respectively, according to
\citealt{burrisetal00}), leading to
$f_{1\,\mathrm{Eu}}=0.97\,f_{\mathrm{total\,Eu}}$ and
$f_{2\,\mathrm{Eu}}=0.03\,f_{\mathrm{total\,Eu}}$. Th is
synthesized exclusively by the r-process and its production was
considered proportional to the production of Eu:
$f_{1\,\mathrm{Th}}=\chi\,f_{\mathrm{total\,Eu}}$ and
$f_{2\,\mathrm{Th}}=0$. The parameter~$\chi$ used was the one that
provided the best simultaneous fit of the [Th/H], [Th/Fe], and
[Th/Eu] vs. [Fe/H] curves, obtained by the model, to our stellar
data. Table~\ref{tab:description_of_GDCE_model_parameters}
provides summarised descriptions of the main GDCE model
parameters, and the final values adopted are listed in
Table~\ref{tab:GDCE_model_parameters}, as a function of Galactic
disk age.

\begin{table}
\caption[]{Summarised descriptions of the main GDCE model
parameters.} \label{tab:description_of_GDCE_model_parameters}
\begin{tabular}{ c l }
\hline \hline
Parameter & Short description\\
\hline
$k$ & Small positive integer used to parametrize the mass\\
& infall rate $F$.\\
$\Delta$ & Time after the start of stellar formation, after which\\
& delayed-production elements start being ejected.\\
$\omega$ & Efficiency of interstellar gas conversion into stars.\\
$u_0$ & Arbitrary parameter in the mass infall rate $F$.\\
$u_1$ & Time of transition between simple and Clayton\\
&model, multiplied by $\omega$.\\
$u_2$ & Time when residue formation becomes efficient,\\
&multiplied by $\omega$.\\
$u_{T_{\mathrm{G}}}$ & Age of the Galactic disk, multiplied by $\omega$.\\
$j$ & IMF slope corresponding to residues that evaporate.\\
$\Gamma$ & Dilution of ISM due to residue evaporation.\\
$f_{1\,\mathrm{Fe}}$ & Fe yield divided by Fe solar abundance today\\
&(prompt contribution).\\
$f_{2\,\mathrm{Fe}}$ & Fe yield divided by Fe solar abundance today\\
&(delayed contribution).\\
$f_{\mathrm{total\,Eu}}$ & Eu yield divided by Eu solar abundance today.\\
$f_{1\,\mathrm{Eu}}$ & Prompt contribution of $f_{\mathrm{total\,Eu}}$.\\
$f_{2\,\mathrm{Eu}}$ & Delayed contribution of $f_{\mathrm{total\,Eu}}$.\\
$f_{1\,\mathrm{Th}}$ & Th yield divided by Th solar abundance today.\\
$\chi$ & Relation between $f_{1\,\mathrm{Th}}$ and $f_{\mathrm{total\,Eu}}$.\\
\hline
\end{tabular}
\end{table}

\begin{table}
\caption[]{Adopted GDCE model parameters. For a short description
of the entries, see
Table~\ref{tab:description_of_GDCE_model_parameters}; details can
be found in the text.} \label{tab:GDCE_model_parameters}
\begin{tabular}{ c c c c c }
\hline \hline
 & \multicolumn{4}{c}{$T_{\mathrm{G}}$}\\
\cline{2-5}Parameter & 6~Gyr & 9~Gyr & 12~Gyr & 15~Gyr\\
\hline
$k$ & \multicolumn{4}{l}{\ {-----------------}\ \ 2\ \ {-------------------}}\\
$\Delta$ (Gyr) & \multicolumn{4}{l}
{{-----------------}\ \ 1.1\ \ {------------------}}\\
$\omega$ & 0.746 & 0.488 & 0.362 & 0.285\\
$u_0$ & 2.040 & 1.500 & 1.250 & 1.110\\
$u_1$ & \multicolumn{4}{l}
{{----------------}\ \ 0.140\ \ {----------------}}\\
$u_2$ & 1.880 & 1.845 & 1.824 & 1.796\\
$u_{T_{\mathrm{G}}}$ & 4.476 & 4.392 & 4.344 & 4.275\\
$j$ & 2.605 & 2.586 & 2.571 & 2.562\\
$\Gamma$ & 0.301 & 0.251 & 0.217 & 0.199\\
$f_{1\,\mathrm{Fe}}$ & \multicolumn{4}{l}
{{----------------}\ \ 0.240\ \ {----------------}}\\
$f_{2\,\mathrm{Fe}}$ & \multicolumn{4}{l}
{{----------------}\ \ 0.467\ \ {----------------}}\\
$f_{\mathrm{total\,Eu}}$ & 0.630 & 0.670 & 0.710 & 0.750\\
$f_{1\,\mathrm{Eu}}$ & 0.611 & 0.650 & 0.689 & 0.728\\
$f_{2\,\mathrm{Eu}}$ & 0.019 & 0.020 & 0.021 & 0.022\\
$f_{1\,\mathrm{Th}}$ & 0.740 & 0.905 & 1.186 & 1.425\\
$\chi$ & 1.175 & 1.350 & 1.670 & 1.900\\
\hline
\end{tabular}
\begin{tabbing}
Observations: \= $u_{T_{\mathrm{G}}}=\omega\,T_{\mathrm{G}}$;
$f_{1\,\mathrm{Eu}}=0.97\,f_{\mathrm{total\,Eu}}$;\\
\> $f_{2\,\mathrm{Eu}}=0.03\,f_{\mathrm{total\,Eu}}$;
$f_{1\,\mathrm{Th}}=\chi\,f_{\mathrm{total\,Eu}}$.
\end{tabbing}
\end{table}

\subsection{The age of the Galactic disk}

We have estimated the age of the Galactic disk in two different
ways: using our GDCE model along with literature production ratio
data, and with our own stellar abundance data.

\subsubsection{$T_{\mathrm{G}}$ from literature data}

Inspection of Table~\ref{tab:GDCE_model_parameters} shows that
parameter~$\chi$ is not constant, but dependent on the Galactic
disk age. Therefore, the age can be estimated by constructing a
$\chi$ vs. $T_{\mathrm{G}}$ diagram, fitting a curve to the
diagram to determine a relation between the two parameters, and
then using literature determinations of $\chi$. Parameter $\chi$
can be calculated from literature determinations of solar Eu and
Th abundances and Th/Eu production ratio through the relation
$\chi=(p^{\mathrm{Th}}/p^{\mathrm{Eu}})/(\mbox{Th}/\mbox{Eu})_{\mbox{\scriptsize\sun}}$,
where $(\mbox{Th}/\mbox{Eu})_{\mbox{\scriptsize\sun}}$ is the
solar abundance ratio today.

From the meteoritic determinations originally presented by
\citet{lodders03} we have
$\log\varepsilon(\mbox{Th})_{\mbox{\scriptsize\sun}}=+0.06\pm0.04$
and
$\log\varepsilon(\mbox{Eu})_{\mbox{\scriptsize\sun}}=+0.49\pm0.04$,
which leads to
$(\mbox{Th}/\mbox{Eu})_{\mbox{\scriptsize\sun}}=+0.372\pm0.069$.
These abundances were not taken directly from \citet{lodders03},
but rather from \citet{asplundetal05}, who corrected their
zero-point in order to make them compatible with their Si
photospheric value. It is much more difficult to estimate the
production ratio $p^{\mathrm{Th}}/p^{\mathrm{Eu}}$. Some of the
works from the literature that estimate the ages of UMP stars
through Th/Eu nucleocosmochronology use the solar Th/Eu abundance
ratio, corrected for the Th decay since the formation of the Sun,
as an estimate of the production ratio. This approach would lead
to a good estimate only if all solar Th were produced in a single
(or at most a few) nucleosynthesis episodes just prior to the
formation of the Solar System. This hypothesis is known to be
false, and stellar ages estimated with production ratios obtained
this way yield lower limits only. For this reason, it is
preferable to use theoretical estimates.

Theoretical production ratios are derived from models of r-process
nucleosynthesis, and subject to many uncertainties, mainly because
the astrophysical site of production by this process is not well
known yet, and because it is not possible to obtain laboratory
data on many very-neutron-rich elements which are very far from
the $\beta$-stability valley. A procedure frequently employed to
reduce these uncertainties is to constrain the model results to
solar abundances. Heavy elements currently present in the Sun have
been synthesized by both the r-process and s-process (with the
exception of U and Th, produced exclusively by the r-process).
R-process nucleosynthesis models are then forced to reproduce just
the fraction of the solar abundances that was produced by the
r-process. This constraint is valid only if all stars produce
r-process elements in solar proportions. The abundances of heavy
elements in UMP stars could also be used to constrain the
r-process nucleosynthesis models, but their high uncertainty makes
it preferable to use the solar abundances.

Many works from the literature \citep[\space among
others]{snedenetal96,snedenetal98,snedenetal00a,snedenetal00b,
westinetal00,burrisetal00,johnson&bolte01,cowanetal99,cowanetal02b}
corroborate the so-called \emph{universality of the r-process
abundances}. According to this hypothesis, r-process
nucleosynthesis would occur at only one astrophysical site and
would always produce elements in solar proportions. These works
support the universality hypothesis by the successful comparison
of heavy element abundances in UMP stars (representing the product
of nucleosynthesis from few preceding stars) with r-process
fraction of solar abundances (representing the cumulative
production during several Gyr of Galactic evolution). This
comparison is carried out for the second and third r-process peaks
only ($56\le\mbox{Z}\le72$ and $73\le\mbox{Z}\le82$,
respectively).\footnote{Some of the works cited above show that
there is no agreement for elements of the first r-process peak
($\mbox{Z}\le55$). This hints at the existence of two different
sites of r-process nucleosynthesis: one that produces low mass
elements (first peak), and one that produces high mass elements
(second and third peaks, and actinides). The non-universality of
low mass elements does not represent a problem for Th/Eu
nucleocosmochronology, since both Eu and Th have atomic numbers
$\mbox{Z}>55$.} To enable the use of r-process nucleosynthesis
models which are constrained to solar abundances in the
calculation of Th/Eu production ratios, Th must also be included
in the universality. It is usually \emph{assumed} that, if the
universality is valid for second and third peaks, it can be
extended to the actinides. Determination of more precise Pb and Bi
abundances in the Sun and UMP stars would allow us to better
investigate if this extension is indeed valid or not. Since these
elements are produced mainly (more than 80\%) by the
$\alpha$-decay of elements with atomic mass $209<\mbox{A}<255$, it
would be very unlikely that solar Pb and Bi abundances would be
compatible with those from UMP stars if universality was not valid
for elements with $\mbox{A}>209$.

The supposition that universality can be extended to the actinides
is still a matter of strong polemic in the literature. In
particular, the analyses of the halo UMP giant {CS~31\,082-001}
carried out by \citet{cayreletal01} and \citet{hilletal02} provide
data \emph{against} it. This star presents a solar abundance
pattern for elements of the second peak, but its third peak does
not match the Sun. While Os and Ir seem to be overabundant
relative to the solar r-process fraction, Pb appears to be
deficient. A more significant discrepancy is that the star has an
abundance ratio $\mbox{Th/Eu}=+0.603$, which is much larger than
the solar value corrected for Th decay
$\mbox{(Th/Eu)}_{\mbox{\scriptsize\sun}}=+0.434$; UMP stars are
expected to present Th/Eu abundance ratios considerably lower than
the solar corrected value, due to their higher age. This high
value indicates that the star could have been formed with an
inherent actinide superabundance relative to the elements of the
second peak, which means that universality would not hold for Th.
{CS~31\,082-001} is the only star hitherto studied that presents
important discrepancies for elements with $\mbox{Z}>55$, whereas
evidence favorable for the validity of universality for these
atomic masses is becoming increasingly numerous. It is not clear
if {CS~31\,082-001} is merely chemically peculiar or if its
discrepancies could be present in other yet unobserved stars; if
the latter is true, it may not be possible to extend universality
to the actinides. If Eu and Th are produced in different
astrophysical sites, the Th/Eu production ratio may have suffered
variations during the evolution of the Galaxy, provoked by changes
in the relative frequencies of their respective progenitors.

\begin{figure}
\resizebox{\hsize}{!}{\includegraphics*{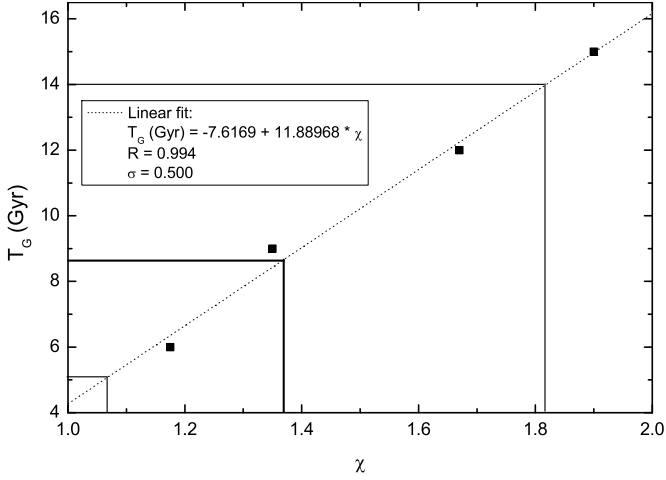}}
\caption{$T_{\mathrm{G}}$ vs. $\chi$ diagram for our GDCE model.
The dotted curve is a linear fit. The solid lines display the
value adopted from the literature for the $\chi$ parameter
($1.371^{+0.444}_{-0.305}$) and its respective disk age
($8.7^{+5.3}_{-3.6}~\mbox{Gyr}$).} \label{fig:chi_vs_age}
\end{figure}

In this paper, we assume that universality \emph{can} be extended
to Th, considering that indications to the contrary are very
sparse. But even if constraining the nucleosynthesis models to
solar abundances reduces the uncertainties, they nevertheless
remain very high, because the calculations can be based on
different far-from-stability mass models and other theoretical
suppositions (e.g., neutron densities).
\citet{cowanetal97,cowanetal99} and \citet{schatzetal02} derive
values between 0.47 and 0.55. We adopted the center of the
interval, $p^{\mathrm{Th}}/p^{\mathrm{Eu}}=0.51\pm0.04$, arriving
at $\chi=1.371^{+0.444}_{-0.305}$. In Fig.~\ref{fig:chi_vs_age}
the $T_{\mathrm{G}}$ vs. $\chi$ diagram constructed for our GDCE
model is presented. A linear fit to the data was used to estimate
a Galactic disk age of
$T_{\mathrm{G}}=8.7^{+5.3}_{-3.6}~\mbox{Gyr}$. Taking into account
the linear fit scatter ($\sigma=0.5$), we arrive at a final value
$T_{\mathrm{G}}=8.7^{+5.8}_{-4.1}~\mbox{Gyr}$. The high
uncertainty of this result reflects the difficulties of estimating
the Th/Eu production and solar abundance ratios, as well as the
low sensitivity of the Galactic disk age to parameter~$\chi$.

\subsubsection{$T_{\mathrm{G}}$ from our stellar data}

\begin{figure}
\resizebox{\hsize}{!}{\includegraphics*{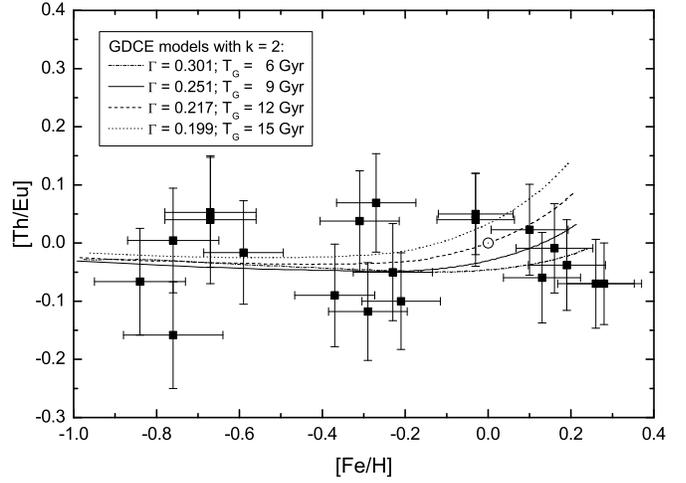}}
\caption{[Th/Eu] vs. [Fe/H] diagram for all sample stars. Curves
were calculated using our GDCE model for the four different
Galactic disk ages shown in the legend.}
\label{fig:th_eu_fe_h_curves}
\end{figure}

[Th/Eu] vs. [Fe/H] curves were constructed from our GDCE model for
each Galactic disk age. These curves, along with our stellar
abundance ratio data, can be seen in
Fig.~\ref{fig:th_eu_fe_h_curves}. Note that the curves draw
further apart at high metallicities. This has a very important
consequence for the analysis: contrary to the intuitive notion
that the most metal-poor stars are the most important, it is the
most metal-rich ones which better discriminate between different
ages. In order to determine the age that best fits our data, we
calculated a quadratic deviation between our data points and each
curve, according to the equation
\begin{equation}
\label{eqn:deviations} \mbox{Total deviation} =
\sum_{i=1}^{19}\left\{[\mbox{Th/Eu}]_i-f([\mbox{Fe/H}]_i)\right\}^2,
\end{equation}
where the index~$i$ represents the {$i$-th} star, and
$f([\mbox{Fe/H}]_i)$ is a $4^{\mathrm{th}}$~order polynomial fit
to the GDCE curve. We were not able to use all 21 sample stars to
calculate the deviations because the two most metal-rich objects
(\object{HD~128\,620} and \object{HD~160\,691}) fall out of the
interval where the curves are defined; that is why the total
deviations were calculated by summing 19 individual stellar
deviations, and not 21.

\begin{figure}
\resizebox{\hsize}{!}{\includegraphics*{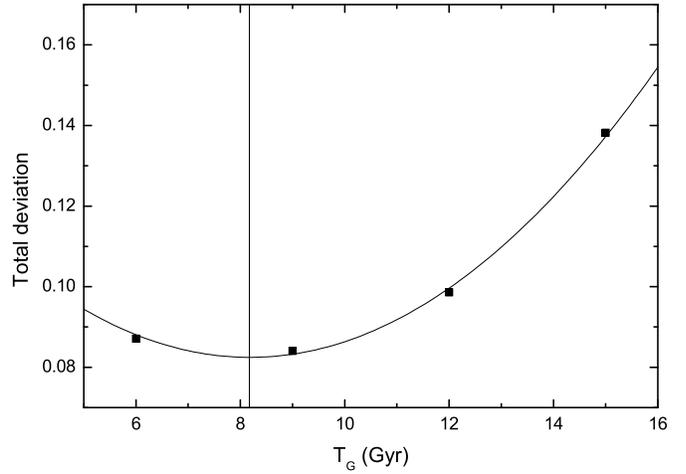}}
\caption{Determination of Galactic disk age that best fits our
stellar [Th/Eu] abundance ratio data. Points are the total
deviations between each curve and the stellar data presented in
Fig.~\ref{fig:th_eu_fe_h_curves}, calculated according to
Equation~\ref{eqn:deviations}. The curve is a
$2^{\mathrm{nd}}$~order polynomial fit. The vertical line marks
the polynomial minimum, corresponding to best Galactic disk age
$T_{\mathrm{G}}=8.2~\mbox{Gyr}$.} \label{fig:deviation_vs_TG}
\end{figure}

Once a deviation was determined for each curve, we traced a
deviation vs. $T_{\mathrm{G}}$ diagram, and fitted a
$2^{\mathrm{nd}}$~order polynomial to it
(Fig.~\ref{fig:deviation_vs_TG}). The Galactic disk age that best
fits our stellar [Th/Eu] abundance ratio data was obtained by
minimising the fitted polynomial: $T_{\mathrm{G}}=8.2~\mbox{Gyr}$.

We estimated the Galactic disk age uncertainty related to the
abundance ratio uncertainties by Monte Carlo simulation. We
developed a code that adds Gaussian random errors to the stellar
[Th/Eu] and [Fe/H] abundance ratios, where the width of the
Gaussians are the uncertainties adopted for the respective
abundance ratio. After this, the code calculates the new total
deviations between the GDCE curves and the modified stellar data,
fits a $2^{\mathrm{nd}}$~order polynomial to the [deviation,
$T_{\mathrm{G}}$] data, and calculates the polynomial minimum,
obtaining a new age. It is possible to choose the number of
simulations, each time with new abundance ratios, added to
different Gaussian errors. The ages obtained are saved in a file
for further analysis.

\begin{figure}
\resizebox{\hsize}{!}{\includegraphics*{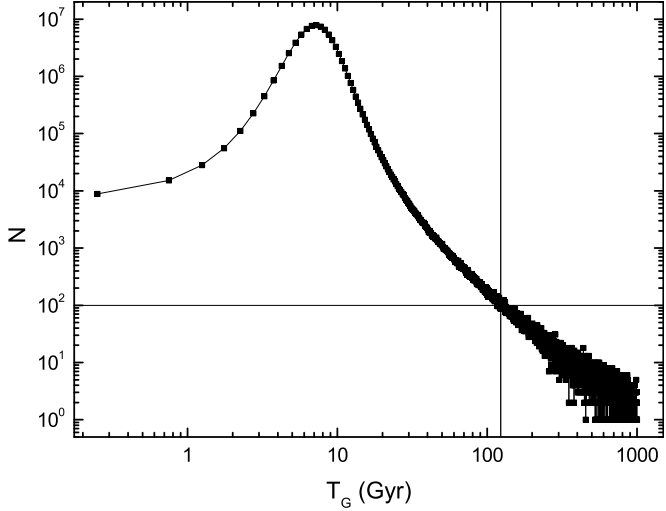}}
\caption{Galactic disk age distribution obtained through a Monte
Carlo simulation of the [Th/Eu] and [Fe/H] abundance ratio
uncertainties. Simulation results have been counted in 0.5~Gyr
bins, and N is the total count per bin. The vertical line marks
the age at which the distribution was cut (125~Gyr), corresponding
to a count fall below 100 (i.e., 10\% uncertainty).}
\label{fig:TG_uncertainties}
\end{figure}

We carried out 75~million simulations. The Galactic disk age
distribution obtained is presented in
Fig.~\ref{fig:TG_uncertainties}, where ages were counted in
0.5~Gyr bins. The few negative ages obtained ($<0.2\%$) were
removed, since they make no physical sense. The distribution has a
clear increase in noise as the age increases (after the mode).
This happens because the distribution was constructed by
\emph{counting} the number of ages between two values. Thus, as
the age increases, the number of counts decreases, and statistical
fluctuation increases. Because of this, we truncated the
distribution at 125~Gyr, where the counts drop below 100
(corresponding to a 10\% statistical uncertainty). One other
issue, besides the noise, forced us to truncate the distribution
at a point where the counts are still statistically significant. A
continuous probability distribution must fall asymptotically to
zero. But this cannot occur with our distribution, as it was
constructed by \emph{counting}, and counts have no fractional
values between 0 and 1. Hence, the result of the calculations
performed to derive an age uncertainty would be incorrect if we
took into consideration the complete distribution.

\begin{figure}
\resizebox{\hsize}{!}{\includegraphics*{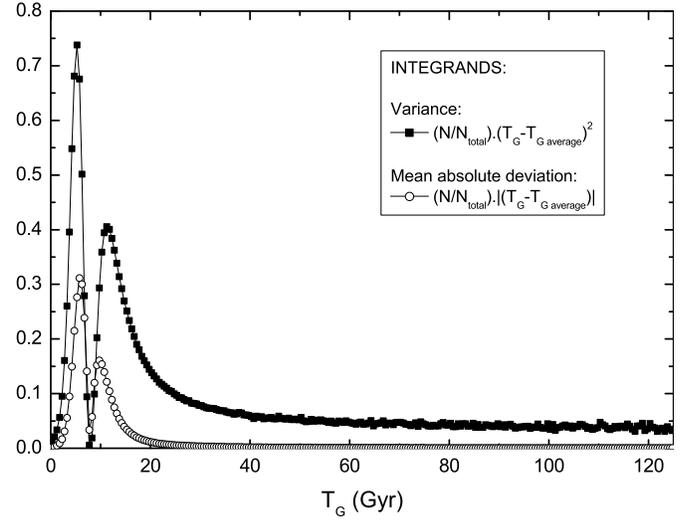}}
\caption{Integrands of Equations~\ref{eqn:variance} and
\ref{eqn:MAbD}, used to calculate the variance and the mean
absolute deviation (MAbD) of the age distribution, respectively.}
\label{fig:variance_and_MAbD}
\end{figure}

A good estimate for our age uncertainty is the width of the
distribution. There is more than one way to estimate this width.
The most widely used is the \emph{variance}, determined through
the equation
\begin{equation}
\label{eqn:variance}
\sigma^2=\int_0^{125}\frac{N}{N_{\mathrm{total}}}(T_{\mathrm{G}}-\overline{T}_{\mathrm{G}})^2\,\mbox{d}T_{\mathrm{G}},
\end{equation}
where $N_{\mathrm{total}}$ is the total number of simulations
within the integration limits and $\overline{T}_{\mathrm{G}}$ is
the average age. The integration limits should have been $-\infty$
and $+\infty$, but since the distribution was truncated at 0 and
125, we carried out the integration accordingly. In
Fig.~\ref{fig:variance_and_MAbD} the integrand of
Equation~\ref{eqn:variance} is presented, and it can be seen that
its right wing falls very slowly. Even if we extended the upper
integration limit to $+\infty$, it is possible that convergence
could not be achieved. Even if we managed to achieve convergence,
the value obtained would be very large, and not representative of
the width of the distribution.

A more robust estimator of the width is the \emph{mean absolute
deviation} (MAbD, see \citealt{pressetal92}):
\begin{equation}
\label{eqn:MAbD} \mbox{MAbD}=
\int_0^{125}\frac{N}{N_{\mathrm{total}}}|T_{\mathrm{G}}-\overline{T}_{\mathrm{G}}|\,\mbox{d}T_{\mathrm{G}}.
\end{equation}
As can be seen in Fig.~\ref{fig:variance_and_MAbD}, the integrand
of Equation~\ref{eqn:MAbD} falls to zero much faster than that of
Equation~\ref{eqn:variance}. Therefore, the integration converges,
and the error introduced by truncating the distribution can be
neglected. Carrying out a numeric integration, we arrive at
$\mbox{MAbD}=1.9~\mbox{Gyr}$. The procedure was repeated several
times, with different bin sizes, and we verified that the MAbD is
not dependent on the bin size.

The final value obtained using our stellar data is
$T_{\mathrm{G}}=(8.2\pm1.9)~\mbox{Gyr}$. It must be noted that the
cited uncertainty is related to the stellar [Th/Eu] and [Fe/H]
abundance ratio uncertainties alone, and does not take into
consideration the uncertainties of the GDCE model itself, which
are very difficult to estimate. The uncertainty related to the
model could very well be the main source of age uncertainty.

\subsubsection{Adopted Galactic thin disk age}

The adopted Galactic disk age was calculated by combining the two
estimates obtained with literature data and with our stellar
abundance ratios. These values were combined using the maximum
likelihood method, assuming that each one follows a Gaussian
probability distribution, which results in a weighted average
using the reciprocal of the square uncertainties as weights. The
final, adopted Galactic disk age is
$$\mbox{FINAL\space}T_{\mathrm{G}}=(8.3\pm1.8)~\mbox{Gyr.}$$

\section{Conclusions}

[Th/Eu] vs. [Fe/H] curves have been constructed for four Galactic
disk ages (6, 9, 12, and 15~Gyr) from a GDCE model developed by
us. These curves were compared to the stellar abundance ratio data
obtained in Paper~I in order to determine the Galactic disk age.
The age that best fits our data was obtained by minimising the
total quadratic deviation between the stellar data and the
theoretical curves. A Monte Carlo simulation was carried out to
estimate the age uncertainty related to the abundance ratio
uncertainties. The value obtained was
$T_{\mathrm{G}}=(8.2\pm1.9)~\mbox{Gyr}$. The age was also
estimated using literature data along with our GDCE model,
yielding $T_{\mathrm{G}}=8.7^{+5.8}_{-4.1}~\mbox{Gyr}$.

Our two age estimates were combined using the maximum likelihood
method, resulting in
$\mbox{FINAL\space}T_{\mathrm{G}}=(8.3\pm1.8)~\mbox{Gyr}$. This
result is the first Galactic disk age determined via Th/Eu
nucleocosmochronology, and is compatible with the most recent
white dwarf ages determined via cooling sequence calculations,
which indicate a low age ($\lesssim10~\mbox{Gyr}$) for the disk
\citep{oswaltetal95,bergeronetal97,leggettetal98,knoxetal99,hansenetal02}.

Determination of the Galactic disk age via [Th/Eu]
nucleocosmochronology was found to be very sensitive to
observational uncertainties. The GDCE model is very insensitive to
the choice of disk age, and the curves derived from the model are
very close when compared to the abundance uncertainties. This
leads to an age uncertainty as high as 1.9~Gyr relative to the
abundance ratio uncertainties alone, not taking into account the
uncertainties intrinsic to the GDCE model itself, which are very
difficult to evaluate. The analysis performed with production
ratios and solar abundances taken from the literature presents an
even higher uncertainty. However, considering that we managed to
reduce the Eu and Th abundance scatters significantly, when
compared to the best data currently available in the literature,
we believe that additional improvements may raise the precision of
the analysis even further. Among the possible enhancements, we can
cite the observation of higher resolution, higher S/N ratio
spectra (preferentially with large sized telescopes),
determination of higher precision atomic parameters (e.g., central
wavelengths of absorption lines), and identification of yet
unknown line blends, like those that forced us to include
artificial Fe lines in the spectral syntheses. If part of the data
scatter is real, and not observational, it will not be possible to
reduce it indefinitely. This may be true if the Th/Eu production
ratio is not constant throughout the Galactic evolution, as
abundance analyses of r-process elements in {CS~31\,082-001} seem
to indicate. Future advancements in r-process nucleosynthesis
models may help solve the issues of universality and constancy of
the production ratio, especially when the sites of production by
this process are finally identified.

%%%%%%%%%%%%%%%%%%%%%%%%%%%%%%%%%%%%%%%%%%%%%%%%%%%%%%%%%%%%%%%%%%%%%%%%%%

\begin{acknowledgements}
This paper is based on the PhD thesis of one of the authors
\citep{delpeloso03}. We thank R. de la Reza and G.F. Porto de
Mello for their contributions to this work. EFP acknowledges
financial support from CAPES/PROAP and FAPERJ/FP (grant
E-26/150.567/2003). LS thanks the CNPq, Brazilian Agency, for the
financial support 453529.0.1 and for the grants 301376/86-7 and
304134-2003.1. Finally, we acknowledge the anonymous referee's
thorough revision of the manuscript, and are grateful for the
comments that helped to greatly enhance the final version of the
work.
\end{acknowledgements}

\bibliographystyle{aa}
\bibliography{referencias}

\end{document}